\documentclass[useAMS,usenatbib,usegraphicx]{mn2e}
\usepackage{epsf,rotating,url}
\usepackage{subfigure,amssymb,amsfonts,amstext,amsgen,amsopn,amsxtra,indentfirst,lscape,times,rotating}
\usepackage{longtable}
\usepackage{graphics}
\usepackage{keyval}
\usepackage{trig}
\usepackage{dcolumn}
\usepackage{bm}
\def\beq{\begin{equation}}
\def\eeq{\end{equation}}
\def\bey{\begin{eqnarray}}
\def\eey{\end{eqnarray}}
\def\msun{M_\odot}

\def\kms{\, {\rm km \, s}^{-1} }

\def\grad{{\bf \nabla}}
\def\prd{Phys. Rev. D}
\def\mnras{MNRAS}
\def\apj{ApJ}

\def\nat{Nature}
\def\apjs{ApJ}
\def\apjl{ApJ}
\def\na{New Astronomy}

\def\aap{A \& A}
\def\aj{AJ}
\def\lcdm{{\Lambda}CDM}

\def\aap{Astron. Astrophys.}

\title{Cosmological simulations in MOND: the cluster scale halo mass function with light sterile neutrinos}

\author[G. W. Angus, A. Diaferio, B. Famaey, K. J. van der Heyden]{G. W. Angus$^{1,2}$\thanks{E-mail: angus.gz@gmail.com}, A. Diaferio$^{3,4}$, B. Famaey$^{5}$ and K. J. van der Heyden$^{2}$ \\ 
$^{1}$Department of Physics and Astrophysics, Vrije Universiteit Brussel, Pleinlaan 2, 1050 Brussels, Belgium \\
$^{2}$Astrophysics, Cosmology \& Gravity Centre, University of Cape Town, Private Bag X3, Rondebosch, 7701, South Africa \\
$^{3}$Dipartimento di Fisica, Universit\`a di Torino, Via P. Giuria 1, I-10125, Torino, Italy \\
$^{4}$Istituto Nazionale di Fisica Nucleare, Sezione di Torino, Via P. Giuria 1, I-10125, Torino, Italy\\
$^{5}$Observatoire Astronomique de Strasbourg, CNRS UMR 7550, France\\}
\begin{document}

\date{\today}
\maketitle
\begin{abstract}
We use our Modified Newtonian Dynamics (MOND) cosmological particle-mesh N-body code to investigate the feasibility of structure formation in a framework involving MOND and light sterile neutrinos in the mass range 11 - 300 $eV$: always assuming that $\Omega_{\nu_s}=0.225$ for $H_o=72 \kms Mpc^{-1}$. We run a suite of simulations with variants on the expansion history, cosmological variation of the MOND acceleration constant, different normalisations of the power spectrum of the initial perturbations and interpolating functions. Using various box sizes, but typically with ones of length 256~$Mpc/h$, we compare our simulated halo mass functions with observed cluster mass functions and show that (i) the sterile neutrino mass must be larger than 30 $eV$ to account for the low mass ($M_{200}<10^{14.6}\msun$) clusters of galaxies in MOND and (ii) regardless of sterile neutrino mass or any of the variations we mentioned above, it is not possible to form the correct number of high mass ($M_{200}>10^{15.1}\msun$) clusters of galaxies: there is always a considerable over production. This means that the ansatz of considering the weak-field limit of MOND together with a component of light sterile neutrinos to form structure from $z \sim 200$ fails. If MOND is the correct description of weak-field gravitational dynamics, it could mean that subtle effects of the additional fields in covariant theories of MOND render the ansatz inaccurate, or that the gravity generated by light sterile neutrinos (or by similar hot dark matter particles) is different from that generated by the baryons.
\end{abstract}


\section{Introduction}
\protect\label{sec:intr}
The standard model of cosmology and specifically how it gives rise to the formation of large scale structure is built upon several well-founded assumptions. We assume there is a period of inflation and, as a general prediction, when that epoch ends there is a scale-free power spectrum of perturbations such that $P_i(k)\propto k^{n_s}$, where $n_s$ is either unity, or very near (within 5\%). These perturbations are present in all the cosmological fluids: baryons, cold dark matter (CDM), neutrinos, photons etc. The perturbations in each fluid grow due to gravity, which is assumed to be described adequately by general relativity. The perturbations grow at different rates on different scales depending on the interplay between the different fluids, the relative contribution of each fluid to the combined energy density and the horizon size. The physics of how the coupled perturbations evolve in the aforementioned fluids is well-tested and supported by measurements of the angular power spectrum of the cosmic microwave background (CMB; \citealt{spergel07,komatsu12,hinshaw12,sievers13}) and buttressed by the observation of baryonic acoustic oscillations (\citealt{eisenstein05,percival10}). 

One feature of the model is that the cosmological density of baryons relative to the critical density is is around 4.5\% and there is a more dominant component of not straight-forwardly luminous, non-baryonic matter with density around 22.5\%. The density of baryons is expected to be around 4.5\% from measurements of the primordial synthesis of light nuclei which were in place long before the acoustic peaks in the cosmic microwave background started taking shape. The necessity for ``dark matter'' from measurements of the cosmic microwave background is not an isolated instance since dynamical measurements of the masses of clusters of galaxies demonstrate there to be roughly the same ratio of dark matter to baryons (e.g. \citealt{McGaugh10} and references therein).

As if this was not sufficient verification, cosmological simulations of high resolution have further solidified the model by confirming a mass function of clusters of galaxies, predicted by theoretical analyses (\citealt{press74,bardeen86,bond91,sheth02}), that well matches the observed cluster mass function (see \citealt{reiprich02}, but also \citealt{rines08} and \citealt{vikhlinin09}). All these separate, cross-matching pieces of evidence lend support to the validity of general relativity as the theory of gravity on large scales, the existence of dark matter and the primordial spectrum of fluctuations. The only significant missing description is how to form galaxies. Despite the difficulty of this task, both numerical simulations (\citealt{scannapieco12,silk12}) and semi-analytical models (\citealt{kauffmann99,bower06,somerville08,benson12}) have shown that the standard model can form galaxies with realistic properties, although some serious discrepancies remain (\citealt{klypin99,moore99,gentile04,mcgaugh05b,peebles10,oh11} and extensively reviewed in \citealt{kroupa10,famaey12}). Even though it remains to be shown how this would work in practice, many believe these discrepancies will be resolved by accurate models of the astrophysical processes.

\begin{figure}
\includegraphics[angle=0,width=8.0cm]{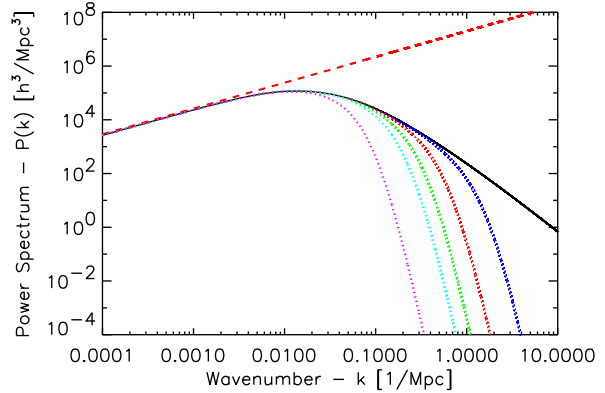}
\caption{Here we plot the linear $z=0$ power spectrum for transfer functions corresponding to CDM (black solid) and 300~eV, 100~eV, 30~eV and 11~eV (blue, red, green, turquoise and magenta dotted lines) sterile neutrinos. The dashed red line is an arbitrarily scaled primordial power spectrum, $P(k)\propto k^{n_s}$.}
\label{fig:tf}
\end{figure}

Returning to the subject of perturbation growth, one can see in Fig~\ref{fig:tf} how the primordial power spectrum of fluctuations is modified on different scales by purely linear growth of perturbations down to $z=0$, depending on whether the dark matter is cold (solid black line) or neutrinos of different mass (dotted lines). The transfer functions (transferring the primordial power spectrum into the late time, matter dominated power spectrum) taken from \cite{abazajian06} are a useful guide until the density contrast of the perturbations with respect to the average density in the universe reaches values close to unity, at which point the analytic models of perturbation growth are not accurate and N-body simulations must be used to follow the non-linear evolution.

Unfolding this evolution in a cosmological setting to the point where galaxies form is a difficult task due to the dynamical scales involved and the complications of hydrodynamical physics. Even in the infancy of galaxy formation studies, it was argued whether the standard cosmological model with CDM and pure general relativity (GR) can produce all the relevant phenomenology of galaxies (see \citealt{kroupa10} for an exhaustive review) with the same success as Milgrom's Modified Newtonian Dynamics (MOND; \citealt{milgrom83a} and see \citealt{famaey12} for a detailed review). MOND is an alternative theory of gravity that modifies Poisson's equation such that the potential responds differently to a mass distribution. In regions of strong gravity, MOND is identical to Newtonian dynamics, but in weak gravity the true gravitational field a test particle experiences from a mass distribution is equal to $\sqrt{\grad\Phi_n a_0}$ - where $\grad\Phi_n$ is the expected Newtonian gravity and $a_0$ is the acceleration constant of the theory.

The longevity of MOND comes from its ability to describe the dynamics of the majority of galactic systems, most remarkably late-type ones where the standard paradigm would need extreme fine-tuning to work. Moreover, it provides a trivial explanation for the independent appearances of the acceleration constant $a_0$ in the zero point of the Tully-Fisher relation, as a transition below which the apparent effect of dark matter rises, or in the observed critical mean surface density for disk stability. However, one cannot ignore the successes of the standard model of cosmology, and MOND should be able to describe the dynamics of clusters of galaxies, the acoustic peaks in the CMB and the formation of the large scale structure from almost homogeneous initial conditions. If MOND is the correct description of galaxy dynamics then there can exist no cosmologically relevant $cold$ dark matter (CDM). Any significant $cold$ dark matter on the galaxy scale would disturb the fits to late-type galaxy rotation curves and the minimal scatter in the baryonic Tully-Fisher relation (see \citealt{McGaugh11,mcgaugh12}).

\cite{nusser02} performed MOND structure simulations with a modified a particle-mesh (PM) code, which included an approximation to the MOND equation. Unfortunately, the simulations were run without dark energy, but generally showed the trend for MOND to produce too much structure on the scales $k=0.1-1.0~k/Mpc$. He showed this conclusion was valid even in an open Universe with $\Omega_m$=0.03 and with an acceleration constant of MOND that is $1/12$ its typical value. \cite{knebe04} incorporated the algebraic MOND equation into their Multi Level Adaptive Particle Mesh (MLAPLM) code and made some similar comparisons to the $\lcdm$ model with MOND as \cite{nusser02}. 

A crucial development on the MLAPM code was made by \cite{llinares08} and further used in \cite{llinares09} because, instead of ``merely" solving the algebraic MOND equation, they solved the momentum and energy conserving modified Poisson equation of the \cite{bekenstein84} theory using the same technique as originally outlined by \cite{brada99a} and further used by \cite{tcevol} for galaxy simulations.

The major problem with these pioneering works was that they lacked a MOND motivated cosmological model. It has been postulated (\citealt{angus09,afd10,ad11} and extensively reviewed in \citealt{da12}) that coupling MOND with a substantial relic abundance of dark matter in the form of sterile neutrinos provides a model of cosmological structure formation that may be able to compete with the well established CDM paradigm. Note that the only difference here is that CDM is traded for a similar abundance of sterile neutrinos and now MOND describes the ultra-weak field accelerations - but still uses GR to describe the cosmological dynamics of the expansion at all times and the growth of perturbations until recombination.

This MOND+sterile neutrino model was motivated primarily by the need for non-baryonic dark matter to explain the relatively high third peak of the cosmic microwave background acoustic power spectrum and the failure of MOND to describe the dynamics of clusters of galaxies (specifically, but not limited to, the bullet cluster; \citealt{clowe04,clowe06,bradac06}).

It was proposed that a sterile neutrino with a mass of $11~eV$ (\citealt{angus09}) could be a promising candidate. It would have to be fully thermalised (one half of all quantum states filled) before freezing out whilst relativistic, since the maximum phase space density for a neutrino of that mass gives $\Omega_{\nu_s}h^2={m_{\nu_s} \over 93.5~eV}=0.118$, which is the same proportion of the critical density as CDM occupies in the CDM model. This was believed to be a natural scenario because if production of these sterile neutrinos was rapid enough then full occupation would be guaranteed, as is the case for the active neutrinos.

Laboratory experiments looking for light sterile neutrinos are one of the most intense research fields in physics. As discussed in the white paper on light sterile neutrinos by \cite{abazajian12}, experiments like the Liquid Scintillator Neutrino Detector (LSND; \citealt{aguilar01}) and MiniBoone (\citealt{maltoni07}) have demonstrated evidence for a fourth (sterile) neutrino with a mass greater than $1~eV$. Although not every other experiment is in agreement with their findings, the expectation still remains - from combined analyses (e.g. \citealt{giunti12}) - that there is a light sterile neutrino around the $1~eV$ level, although this evidence remains weak and by no means conclusive. The masses of sterile neutrinos we are interested are significantly more massive than this, somewhere in the range of 11 - 300~eV, and are currently only constrained in the standard context by model dependent phase space arguments related to the formation of dwarf galaxy dark matter halos (\citealt{boyarsky09}) and the clustering of Lyman-$\alpha$ clouds (\citealt{seljak06,viel06}). The power spectrum of Lyman-$\alpha$ absorbers at high redshifts currently rules out any sterile neutrino particle with a mass less than $\sim 2~keV$ (\citealt{boyarsky09b}) because such a particle would suppress power on $k>1~Mpc/h$ scales. In our model, the clustering of these clouds would not be due to the gravitational attraction towards small scale dark matter halos, which do not exist in MOND, but towards the purely baryonic galaxies required in MOND. Therefore, like with galaxy formation, we must wait for MOND hydrodynamical simulations in a self-consistent cosmology to test our adherence to these important observations.

This MOND plus $11~eV$ sterile neutrino model is clearly not a standard warm dark matter model and had the potential to fulfill the two aforementioned gaps in the cosmology of MOND (the CMB and dark matter in clusters), the most pressing question was whether it could go on to produce the correct distribution of large scale structure in the universe. To this end, \cite{ad11} (hereafter AD11) developed a particle-mesh Poisson solver to perform cosmological simulations in the framework of Quasi-linear MOND (QUMOND; \citealt{milgrom10}). QUMOND and AQUAL (\citealt{bekenstein84}) are both classical theories of gravity, with identical phenomenologies, derived from an action and obey the relevant conservation laws. QUMOND was chosen over AQUAL because it is easier to work with numerically (\citealt{llinares08,llinares09}).

Since there is no widely accepted covariant version of MOND (e.g. \citealt{bekenstein04}, but see \citealt{skordis09,famaey12} for reviews), it was assumed that general relativity (GR) is the correct description of gravity during the radiation dominated phase of the universe. This allows us to use the GR results for the growth of perturbations until we start our simulations around $z\sim200$. For this to be the case, we would require the acceleration constant of MOND to be zero at redshifts $z>200$. Our other assumption is that the expansion history of the universe follows the Friedmann-Robertson-Walker (FRW) rate for $\Omega_M=0.27$ and $\Omega_{\Lambda}=0.73$. A more comprehensive theory should provide its own Friedmann like expansion history and a proper mechanism for the changing of the acceleration constant with redshift, but we do not investigate these here.

The simulations in AD11 used boxes of length $512~Mpc/h$, 256 cells and particles per dimension. As a result, the mass and spatial resolution was extremely coarse. This made it impossible to determine if the simulations produced the correct halo mass function at the scale of clusters of galaxies, but nevertheless, it was clear that there was a catastrophic over-production of supercluster sized halos and large voids (with 250~Mpc/h diameter) were also a hallmark. The apparent lack of production of cluster sized halos in the $11~eV$ sterile neutrino simulations begs the question of whether higher mass neutrinos could perform more adequately.

There is no necessity for the sterile neutrino to weigh $11~eV$. This is the lower limit to have a sufficiently high phase space density to permit the large dark matter densities in MONDian low mass clusters and groups of galaxies. The upper limit, like the lower limit, is set by the free streaming scale because we cannot have sterile neutrino halos being too compact since they would interfere with the typically excellent fits to galaxy rotation curves found with MOND. This is difficult to judge without cosmological simulations that resolve the formation of these smallest of halos, but the free-streaming scale is roughly $L_{fs}=\left({10~eV \over m_{\nu_s}}\right)~Mpc/h$. If we wish to avoid halos on scales smaller than the typical extent of large disks (say 50~kpc), then an upper mass limit of $m_{\nu_s}<300~eV$ should be applied. One additional salient feature of a more massive sterile neutrino is that it is not fully thermalised (only a tiny fraction of the quantum states are occupied as is the case for CDM) and as such would have a significantly lower energy density whilst relativistic than a fully thermalised species. This would positively influence the relic abundance of $^4He$ which is overproduced with 4 thermalised neutrino species.

Whether this over-production of superclusters is a result of resolution or neutrino mass is important to understand. In this paper we compare the observed mass function of clusters of galaxies with our simulated halo mass functions: using a spectrum of neutrino masses, MOND interpolating functions and cosmological dependence on the acceleration constant of MOND to ultimately decide whether a cosmological model using MOND plus massive sterile neutrinos can possibly reproduce the large scale structure of our universe.
\section{METHODS}
\protect\label{sec:code}

\subsection{QUMOND}

The Aquadratic Lagrangian theory of \cite{bekenstein84} produces a modified Poisson equation that must be solved numerically for arbitrary geometries. Likewise, QUMOND (\citealt{milgrom10}) requires solution of a modified Poisson equation, but one that is slightly easier to implement. Specifically, the ordinary Poisson equation for cosmological simulations

\beq
\protect\label{eqn:qumond1}
\nabla^2\Phi_N=4\pi G (\rho - \bar{\rho})/a
\eeq
is solved to give the Newtonian potential, $\Phi_N$, at scale factor $a$, from the ordinary matter density $\rho$ that includes baryons and neutrinos. This would also include cold dark matter if there was any in our model. The QUMOND potential, $\Phi$, is found from the Newtonian potential as follows
\beq
\protect\label{eqn:qumond2}
\nabla^2\Phi=\grad \cdot \left[ \nu(y) \grad\Phi_N \right],
\eeq
where $\nu(y)=0.5+0.5\sqrt{1+4/y}$ and $y=\nabla\Phi_N/a_o a$. $a_o$ is the MOND acceleration constant, chosen here to be $3.6~(\kms)^2pc^{-1}$, which is set by fitting the baryonic Tully-Fisher relation. This acceleration constant cannot be adjusted to allow the introduction of dark matter to galaxies because any significant dark matter on galaxy scales would alter the predicted rotation curves and render them incompatible with the measured ones. Adding a component of dark matter to galaxies would also increase scatter to the Baryonic Tully-Fisher relation, which does not appear to be present in the data (\citealt{mcgaugh05b}).

The specifics of how to solve Eqs~\ref{eqn:qumond1} \& \ref{eqn:qumond2} are also explained in AD11, but we review the main points here. The code we use is particle-mesh based. The grid-mesh has $257$ cells in each dimension and we typically use $256$ particles per dimension. The reason we take a particle-mesh approach is that the MOND Poisson equation is non-linear and therefore, we cannot use a direct or tree-code approach because co-adding the individual gravity of many particles would yield incorrect results. We must solve the full MOND Poisson equation because using the MOND equation of spherical symmetry does not respect the conservation laws.

The particle positions and velocities are converted to simulation units at the start of the simulation, following the prescription of \cite{pmcode}. The density of the particles is assigned to the various cells with the cubic cloud-in-cell method. Typically, MOND simulations are plagued by the difficulty of handling the boundary conditions, but the periodic boundary conditions used in cosmological simulations allow this to be easily handled. Once the density has been assigned, multigrid methods (see Numerical Recipes \S19.6) are used with finite differencing techniques to solve the Poisson equation to find the Newtonian potential (Eq~\ref{eqn:qumond1}). In the multigrid methods, we use a 3D black-red sweep to update the cells with the new approximation of the potential in that cell and we iterate until we have fractional accuracy of $10^{-8}$.  Once we have the Newtonian potential in each cell, we take the divergence of the vector in the square brackets of Eq~\ref{eqn:qumond2} which gives us the source of the MOND potential. We then repeat the Poisson solving step with the new source density to give the MOND potential, $\Phi$, which we take the gradient of to find the gravity at each cell. We then interpolate to each particle's position to find the appropriate gravity and move each particle with a second order leapfrog.

\subsection{Initial Conditions}
\protect\label{sec:ics}
We make use of the original \verb'COSMICS/GRAFICS' package of \cite{bert95} to generate our initial conditions. We chose to input our own transfer functions using the massive neutrino parameterisation of \cite{abazajian06} (their equations 10-12) and the resulting linear matter power spectra are plotted for the neutrino masses we used in Fig~\ref{fig:tf}. The initial conditions are produced from these transfer functions and we do not add thermal velocities (see e.g. \citealt{klypin93}). We always use the default combination of cosmological parameters ($\Omega_b$, $\Omega_{\nu_s}$, $\Omega_{\Lambda}$, $h$, $n_s$, $Q_{rms-PS}$)=(0.045, 0.225, 0.73, 0.72, 0.95, $17~\mu K$) unless otherwise stated. The CMB quadrupole ($Q_{rms-PS}$; as discussed in AD11) is used to normalise the initial power spectrum of perturbations in the same way as $\sigma_8$ typically is for CDM simulations, because one cannot use linear theory in MOND to estimate $\sigma_8$ at $z = 0$.

\subsection{Halo Finding}
We performed a series of simulations (listed in table~\ref{tab:sims}) using different combinations of available parameters. All simulations use 256 cells and particles per dimension. We obtain our halo mass function thanks to the Amiga Halo Finder (AHF) of \cite{gill04,knollmann09} which we verified by visual inspection to find all relevant halos and subhalos. AHF provides the spherically averaged density profiles of all the halos in our simulation down to an average enclosed density that is 200 times the critical density, signified by the radius $r_{200}$. We discount all halos found with less than 300 particles. For the Newtonian simulations we find the enclosed mass of particles, $M_p(r)$, at each radius by multiplying the number of enclosed particles by the individual particle mass, $m=1.4\times10^{11}(\Omega_{b}+\Omega_{\nu_s})(L_{box}/N_p)^3\msun$, where $L_{box}$ is the length of the box in $Mpc$ (not $Mpc/h$) and $N_p$ is the number of particles in 1D. In Newtonian gravity, $M_p(r)$ is precisely the Newtonian halo mass $M_n(r)$. For the MOND simulations, we first find the enclosed mass of particles (as per the Newtonian simulations, $M_p(r)$), and use the MOND formula to give $M_m(r)=\nu\left({GM_p(r) \over r^2a_o}\right)M_p(r)$, which is the equivalent Newtonian halo mass for the distribution of particles with a MOND gravitational field. To clarify, $M_m(r)$ is the dynamical mass that would be derived using a dynamical test, assuming Newtonian dynamics, for the given mass distribution if MOND is actually the correct description of gravity. Therefore, $M_m(r)$ and $M_n(r)$ are directly comparable to Newtonian measures of the dynamical mass in clusters of galaxies i.e. the cluster mass functions.

For both $M_m(r)$ and $M_n(r)$ we must calculate the mass enclosed at the radius where the average enclosed density is 200 times the critical density of the universe. This is done by interpolating through the mass profiles of each halo.

\subsection{Comparison with theoretical halo mass function}

In Figs~\ref{fig:300n} and \ref{fig:300m} we plot the Newtonian and MOND halo mass functions for an $m_{\nu_s}=300~eV$ sterile neutrino using a series of box sizes to demonstrate the probable range of suitability of our simulations. For both sets of simulations we used box sizes of 64, 128 and 256~$Mpc/h$ and require each mass bin of 0.23 dex to have five halos. The number of halos per bin decreases towards higher masses and so this means we do not plot the mass function above some mass where the number of halos in that bin is less than five. At the low mass end, the spatial and mass resolution begins to curtail the formation of low mass halos. We know theoretically that the mass function should continue to rise for progressively lower mass halos until the free streaming scale inhibits the formation of any lower mass halos. Therefore, we do not plot the mass function for halo masses lower than where the mass function stops rising - which, as stated above, is mainly due to insufficient spatial resolution. Free streaming is never a problem for our default $m_{\nu_s}=300~eV$ simulations.

The colour of each line in Figs~\ref{fig:300n} and \ref{fig:300m} defines a box size and the dashed and solid lines reflect two randomly different sets of initial conditions for simulations with 300~eV sterile neutrinos. For the Newtonian simulations in Figs~\ref{fig:300n} we also plot CDM simulations with 128 and 256~Mpc/h box sizes. For comparison with the Newtonian simulations, we plot the theoretical halo mass function, using the code of \cite{reed07}, for a $\lcdm$ simulation with $\sigma_8$=0.8 and our default dark matter, baryon fractions and Hubble constant. In the Newtonian simulations, the theoretical cluster mass function is underpredicted by a factor of 2-3 by the 256~Mpc/h boxes for halo masses between $10^{14}$ and $10^{14.9}\msun$, where the spatial resolution is insufficient to form halos to the theoretical limit. Using 128~Mpc/h boxes, the theoretical mass function is quite well matched by the simulations between $10^{13.5}$ and $10^{14.6}\msun$. The 64~Mpc/h box is not as useful because of the low absolute number of halos formed.

These aforementioned simulations all use $256^3$ particles and 257 cells per dimension - only the box size varies. To demonstrate that it is spatial resolution that prevents agreement with the theoretical halo mass function, we plot two further simulated mass functions using $128^3$ particles and 129 cells per dimension for a 128~Mpc/h box using turquoise coloured lines. One can see that they both trace the 256~Mpc/h boxes for low masses until roughly $10^{14.5}\msun$, where small numbers of halos makes it unreliable. Had the spatial resolution not been a problem, then these two lines would have traced the 128~Mpc/h boxes with 257 cells per dimension.

For the MOND simulations there is no theoretical halo mass function to compare with, but we plot the theoretical Newtonian halo mass function for comparison. As per the Newtonian simulations, the 128~$Mpc/h$ box has a slightly higher amplitude than the 256~$Mpc/h$ box. They both have a similar shape, which is very flat with increasing mass and this highlights the strong disagreement of the halo mass function of MOND with the Newtonian model. The reason we did not run simulations on different scales for the 11~eV sterile neutrinos in AD11 is that the 128~$Mpc/h$ boxes are dominated by a single massive halo and no other halos are resolved and the 256~$Mpc/h$ boxes are only slightly better.

On the topic of the halo mass range of suitability, there are pros and cons of using larger or smaller boxes. With smaller boxes the spatial and mass resolution increases, but at the expense of larger statistics from more halos that the larger boxes provide. So the reason the smaller boxes look like they have larger normalisations is that they have the resolution to assist the formation of borderline halos and each halo is therefore more massive than if it had been traced with poorer resolution. This is quite clearly demonstrated with the turquoise line of Fig~\ref{fig:300n} which shows two simulations with different box sizes, but the same spatial resolution have very similar mass functions. In both the Newtonian and MOND simulations, the trend for all three lines (blue, red and black) is quite clear, but the difference between normalisation is significant - especially for the Newtonian simulations. So our assertion is that it is acceptable to use any of the boxes over their given plotted range, but that the normalisation will be inaccurate due to our lack of convergence. Furthermore, as can be seen in Fig~\ref{fig:300m}, the issue with the MOND halo mass functions is an over-production of halos with high mass and therefore our lack of spatial resolution means the true mass function will be slightly larger than we find.

Looking specifically at the MOND simulations it is apparent that only the 256~Mpc/h simulations can properly model the mass function at masses larger than $10^{15}~\msun$, because the statistics of the smaller box simulations is too poor.

With warm dark matter simulations there can be the problem of spurious halos growing on small scales where there is no physical power, only shot noise from the initial conditions (see \citealt{wang07} and more recently \citealt{angulo13}). On the scales we are considering, 1 to 256~Mpc/h, this is not an issue especially for our 300~eV simulations for which the transfer functions only significantly differ from CDM on scales smaller than 1Mpc.

\begin{table*}
\begin{tabular}{|l|cccccc|}
\hline
Name		&MOND or Newton &Length of box	&$Q_{rms-PS}$ or $\sigma_8$	&$m_{\nu_s}$	&$a_M$	&$\nu$-function\\
		&M or N		&$Mpc/h$	&$\mu K$ or $-$ 		&$eV$		&	&\\
\hline
\hline
nu10m		&M		&256		&17				&10		&0	&$\alpha=1$\\
nu30m		&M		&256		&17				&30		&0	&$\alpha=1$\\
nu50m		&M		&256		&17				&50		&0	&$\alpha=1$\\
nu100m		&M		&256		&17				&100		&0	&$\alpha=1$\\
nu300m		&M		&256		&17				&300		&0	&$\alpha=1$\\
nu300m2		&M		&256		&17				&300		&0	&$\alpha=1$\\
cdm08m		&M		&256		&0.8				&cdm		&0	&$\alpha=1$\\
\hline\hline
cdm08n128	&N		&128		&0.8				&cdm		&-	&-\\
cdm08n		&N		&256		&0.8				&cdm		&-	&-\\
cdm08n1		&N		&256		&0.8				&cdm		&-	&-\\
cdm09n		&N		&256		&0.9				&cdm		&-	&-\\
Q-04n		&N		&256		&04				&300		&-	&-\\
Q-07n		&N		&256		&07				&300		&-	&-\\
Q-10n		&N		&256		&10				&300		&-	&-\\
Q-13n		&N		&256		&13				&300		&-	&-\\
Q-16n		&N		&256		&16				&300		&-	&-\\
Q-18n		&N		&256		&18				&300		&-	&-\\
\hline
Q-04m		&M		&256		&04				&300		&0	&$\alpha=1$\\
Q-07m		&M		&256		&07				&300		&0	&$\alpha=1$\\
Q-10m		&M		&256		&10				&300		&0	&$\alpha=1$\\
Q-13m		&M		&256		&13				&300		&0	&$\alpha=1$\\
Q-17m (nu300m)	&M		&256		&17				&300		&0	&$\alpha=1$\\
Q-17am		&M		&256		&17				&300		&0	&$\alpha=1$\\
\hline\hline
a0z1m		&M		&256		&17				&300		&0.1	&$\alpha=1$\\
a0z2m		&M		&256		&17				&300		&0.2	&$\alpha=1$\\
a0z3m		&M		&256		&17				&300		&0.25	&$\alpha=1$\\
a0z4m		&M		&256		&17				&300		&0.3	&$\alpha=1$\\
a0z5m		&M		&256		&17				&300		&0.35	&$\alpha=1$\\
a0z6m		&M		&256		&17				&300		&0.4	&$\alpha=1$\\
a0z7m		&M		&256		&17				&300		&0.5	&$\alpha=1$\\
\hline\hline
b32m		&M		&32		&17				&300		&0	&$\alpha=1$\\
b32am		&M		&32		&17				&300		&0	&$\alpha=1$\\
b64m		&M		&64		&17				&300		&0	&$\alpha=1$\\
b64am		&M		&64		&17				&300		&0	&$\alpha=1$\\
b128m		&M		&128		&17				&300		&0	&$\alpha=1$\\
b128am		&M		&128		&17				&300		&0	&$\alpha=1$\\
\hline
b32n		&N		&32		&17				&300		&0	&$\alpha=1$\\
b32an		&N		&32		&17				&300		&0	&$\alpha=1$\\
b64n		&N		&64		&17				&300		&0	&$\alpha=1$\\
b64an		&N		&64		&17				&300		&0	&$\alpha=1$\\
b128n		&N		&128		&17				&300		&0	&$\alpha=1$\\
b128an		&N		&128		&17				&300		&0	&$\alpha=1$\\
b128c129n	&N		&128 (129 cells)&17				&300		&0	&$\alpha=1$\\
b128c129ncdm	&N		&128 (129 cells)&-0.8				&cdm		&0	&$\alpha=1$\\
\hline\hline
bet-02		&M		&256		&17				&300		&0	&$\beta=0.2$\\
bet-05		&M		&256		&17				&300		&0	&$\beta=0.5$\\
alp-2		&M		&256		&17				&300		&0	&$\alpha=2$\\
\hline\hline
wz50		&M		&256		&17				&300		&0 (w=-0.5)	&$\alpha=1$\\
wz75		&M		&256		&17				&300		&0 (w=-0.75)	&$\alpha=1$\\

\hline\hline
\end{tabular}
\caption{In this table we give the details of all the simulations we ran. The columns have the following entries: (1) Simulation name. (2) Whether the simulation was run using MOND (M) or Newtonian (N) gravity or Modified Baryonic Dynamics (MBD). (3) The 1D length of the box in Mpc/h. (4) The normalisation of the simulation initial conditions, either by the CMB quadrupole (in units of $\mu K$), or $\sigma_8$. (5) The mass of sterile neutrino used in $eV$ (cdm is also used if the initial conditions were for cold dark matter). (6) The scale factor, $a$, at which MOND switches on. (7) The specifics of the $\nu$-function for MOND, either using the parameterisation of Eq~\ref{eqn:nubet} or \ref{eqn:nualp}. }
\protect\label{tab:sims}
\end{table*}

\section{Galaxy cluster mass functions}
\protect\label{sec:massfunc}

\begin{figure}
\includegraphics[angle=0,width=8.0cm]{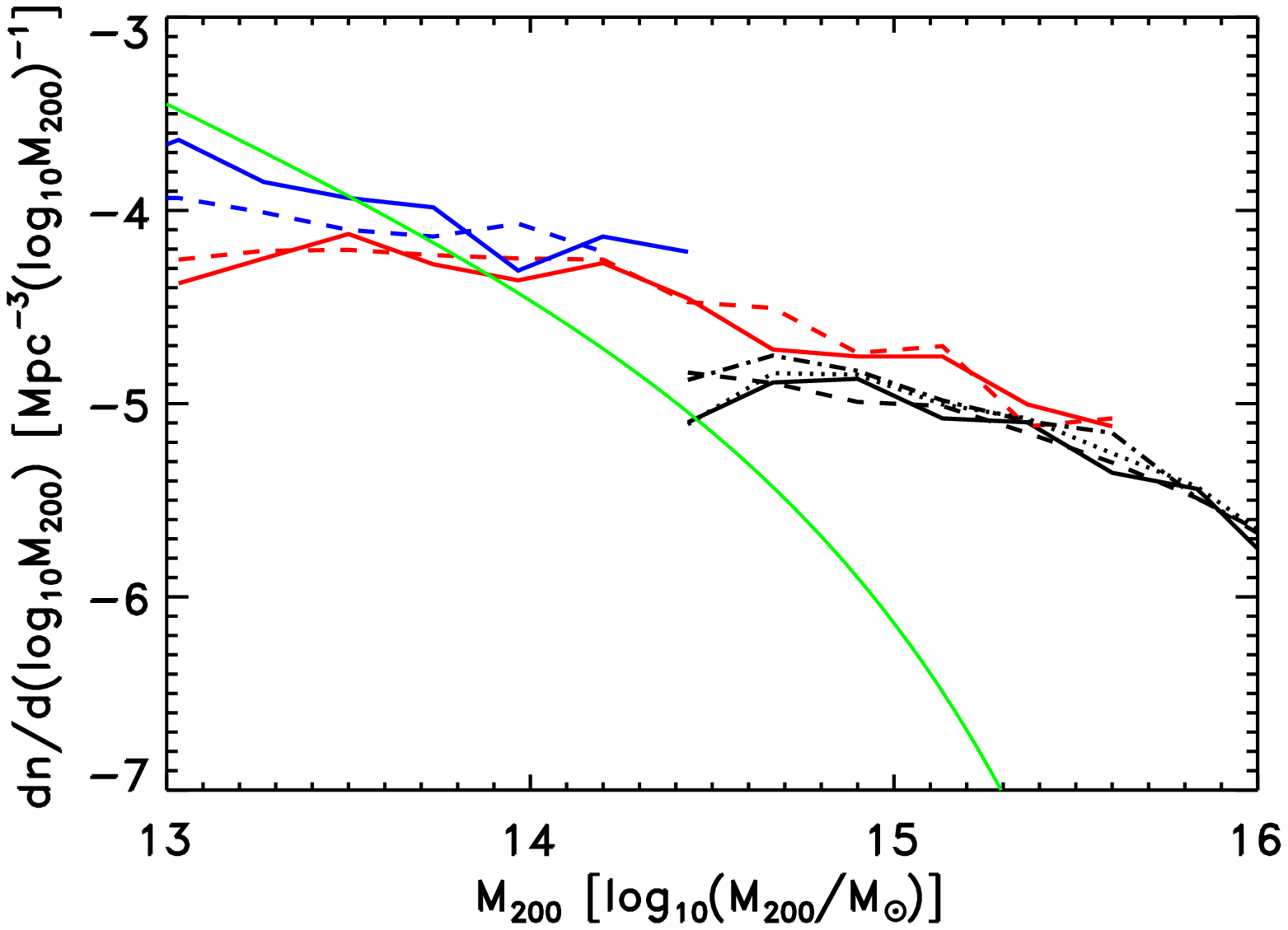}
\caption{Here we plot the MOND halo mass function of a $300~eV$ sterile neutrino using a series of box sizes. The blue, red and black coloured lines correspond to 64, 128, 256~$Mpc/h$ box lengths. The different line-types signify a different realisation of the initial conditions with the same box size. In table 1 these correspond to simulations b64m-b128am (b*m and b*am use solid and dashed line-types respectively), nu300m (solid) and nu300m2 (dashed). We also plot two simulations using 256~$Mpc/h$ boxes, a 300~eV mass and different $\nu$ functions ($\beta=0.2$ and 0.5 using dotted and dot-dashed lines (Table 1:bet-02 and bet-05).}
\label{fig:300m}
\end{figure}

\begin{figure}
\includegraphics[angle=0,width=8.0cm]{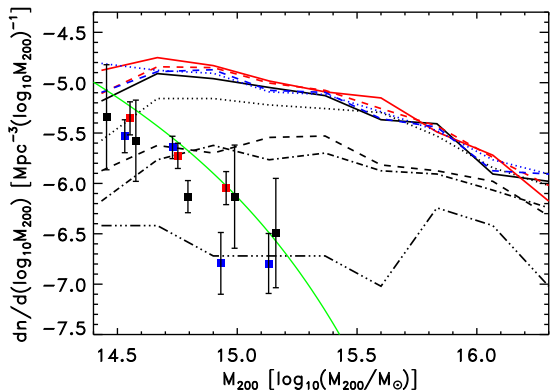}
\caption{Here we plot the MOND halo mass functions for a set of sterile neutrinos of different mass: 11~eV, 30~eV, 50~eV, 100~eV and 300~eV (black triple dot-dashed, dot-dashed, dotted, dashed and solid. Table 1: nu11m-nu300m). We also plot three simulations using a 300~eV mass and different $\nu$ functions ($\beta=0.2$ and 0.5 using red dashed and solid lines respectively and $\alpha=2$ with the dashed blue line - Table 1:bet-02, bet-05 and alp-2) and a simulation with CDM initial conditions that is evolved with MOND with the dotted blue line (Table 1:cdm08-256m). The data points come from RB02 (circles) and RDN08 (squares and triangles found using the virial theorem and the caustic technique respectively). The simulated mass functions are less reliable below 14.7.}
\label{fig:mnu}
\end{figure}

In Fig~\ref{fig:mnu} we plot the mass function of halos in MOND for a series of masses of sterile neutrinos (11~eV to 300~eV) with a 256~Mpc/h box. We compare our halo mass function with the cluster mass functions presented in \cite{reiprich02} and \cite{rines08}, hereafter RB02 and RDN08 respectively. It is obvious that sterile neutrino masses less than or equal to 30~eV are incapable of producing the correct number density of clusters with mass less than $10^{14.6}~\msun$ and more noteworthy, using sterile neutrinos with a mass larger than $30~eV$ precludes forming the correct number density of higher mass clusters $> 10^{14.6}~\msun$. For the observed cluster mass function there is a steep drop off in the number density of clusters near $10^{15}~\msun$, and clearly the predicted number density of MOND halos with mass $10^{15.1}~\msun$ (assuming $m_{\nu_s} > 30~eV$) is between one and two orders of magnitude larger (cf. Figs~\ref{fig:300m} and \ref{fig:mnu}).

To expand on this point, from Fig~\ref{fig:mnu} one can see that increasing the sterile neutrino mass leads to a larger amplitude for the mass function. Therefore, taking our spatial resolution into account, we expect that we can rule out any sterile neutrino mass that yields a mass function lower than the point at $10^{14.6}~\msun$. For a sterile neutrino mass $m_{\nu_s} \le 30~eV$, the $z=0$ mass function is significantly lower than observed for halos with mass less than $10^{14.6}~\msun$ (see Fig~\ref{fig:mnu}). Mergers are not responsible for the eradication of these low mass halos in $m_{\nu_s} = 30~eV$ simulations. This means that if $m_{\nu_s} < 30~eV$ it is not possible to form the correct number of low mass halos. To create the enough halos weighing less than $10^{14.6}~\msun$,  $m_{\nu_s}$ must be greater than $30~eV$.

\begin{figure}
\includegraphics[angle=0,width=8.0cm]{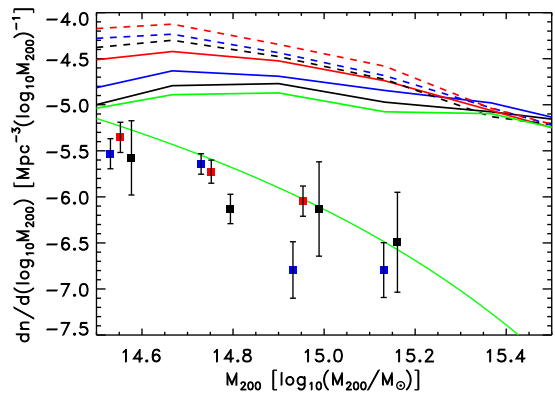}
\caption{Here we plot the halo mass function for a set of $300~eV$ sterile neutrino simulations where MOND is initially switched off and is switched on at a specific scale-factor. The line-types solid green, black, blue and red and dashed black, blue and red correspond to switch on scale-factors of 0, 0.1, 0.2, 0.25, 0.3, 0.4 and 0.5. For simulation details see table 1 and simulation names nu300m and a0z1m-a0z7m. The data points are described in Fig~\ref{fig:mnu}. The simulated mass functions are less reliable below 14.7.}
\label{fig:a0z}
\end{figure}

The question we wish to answer is whether it is possible to use a higher mass sterile neutrino to produce the lower mass halos and impede the formation of the higher mass halos. Our first case was if the acceleration constant of MOND, $a_o$, was smaller at higher redshifts. To test this we ran several simulations where $a_o=0$ until a specific scale-factor, $a$, where the MOND acceleration constant instantaneously took on the standard value $a_o=3.6~(\kms)^2pc^{-1}$. The eight scale-factors we chose were $a=0.0$, 0.1, 0.2, 0.25, 0.3, 0.35, 0.4 and 0.5 and we plot the redshift zero halo mass function for each of these in Fig~\ref{fig:a0z}.

The later MOND switches on ($a=0.5$ - red dashed line - being the latest) the more we can increase the number density of low mass halos relative to the high mass halos, but the absolute number density of high mass halos remains very mildly affected. The earlier MOND switches on ($a=0$ or 0.1 - black solid line - being the earliest) the closer the relative abundance of low mass to high mass halos is and the higher the absolute number density of the high mass halos. The reason for this is that when MOND is switched on, there is a higher prevalence of mergers, which reduces the number of low mass halos, but only mildly affects the masses of the more massive halos. 

There is another factor to consider here and that is galaxies in MOND must form without the aid of a dark matter halo (cold, warm or hot) and galaxy formation without dark matter (if it is possible at all) is only possible with the added benefit of stronger than Newtonian gravitational attraction between the baryons. Thus, if MOND was not in effect until $z=1$, then galaxies would not {\it begin} to form until then and galaxies are clearly formed long before this.

\begin{figure}
\includegraphics[angle=0,width=8.0cm]{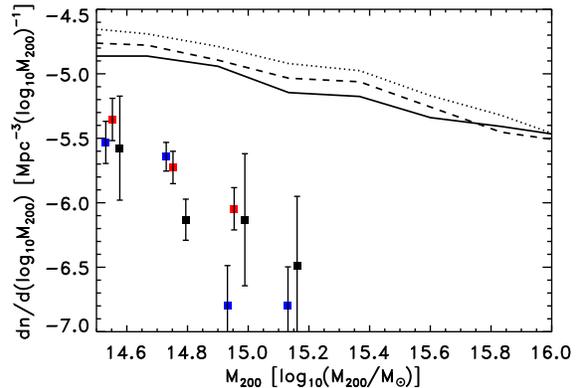}
\caption{Here we plot 3 MOND simulations with the same initial conditions using a 256~Mpc/h box. The three simulations each use a different expansion history according to $\rho_{\Lambda}\propto a^{-3(w+1)}$. The solid line uses $w=-1$, the dashed line uses $w=-0.75$ and the dotted line uses $w=-0.5$ (Table 1:Q-17am, wz75 and wz50). The data points are described in Fig~\ref{fig:mnu}.}
\label{fig:wz}
\end{figure}

We also looked at various different expansion histories with parameterisations of the equation of state of dark energy. In this set up, the energy density of dark energy was re-expressed as $\rho_{\Lambda}\propto a^{-3(w+1)}$. For all other simulations we used the standard $w=-1$, but for two 256 Mpc/h box simulations we used w=-0.75 and w=-0.5. In Fig~\ref{fig:wz} one can see that they are not conducive to subduing the formation of very massive halos and in fact increase the amplitude of the MOND halo mass function.

\begin{figure}
\includegraphics[angle=0,width=8.0cm]{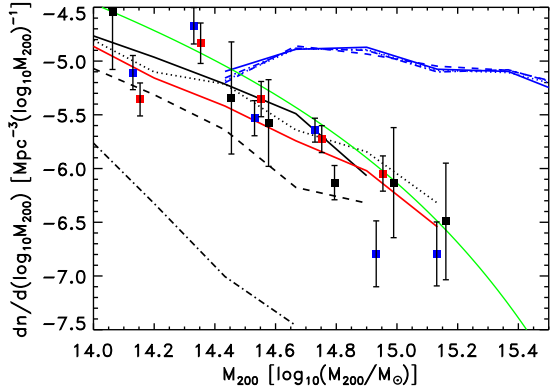}
\caption{The black lines are the halo mass functions for a set of Newtonian $300~eV$ sterile neutrino simulations with various initial perturbation normalisations through the CMB quadrupole $Q_{rms-PS}$. The black line-types dot-dashed, dashed, dotted and solid correspond to $Q_{rms-PS}=10$, 13, 16 and 18~$\mu K$ (table 1:Q-10n, Q-13n, Q-16n and Q-18n respectively). The red line is for CDM initial conditions normalised by $\sigma_8=0.8$ (table 1: cdm08n). The blue lines are partly the same $300~eV$ initial conditions as the black lines correspond to except they are evolved with MOND gravity. Also plotted are two other $300~eV$ simulations using MOND but with different $\nu$ functions. We do not describe these MOND simulations further given the obvious degeneracy but in table 1 they are: Q-10m, Q-13m, Q-17m, bet-02 and bet-05. The data points are described in Fig~\ref{fig:mnu}.}
\label{fig:quad}
\end{figure}

Another case we considered was the normalisation of the initial conditions through the quadrupole of the CMB. We ran six Newtonian simulations with $Q_{rms-PS}=4$, 7, 10, 13, 16 and 18~$\mu K$ for $m_{\nu_s}=300~eV$. In Fig~\ref{fig:quad} one can see the halo mass functions for $Q_{rms-PS}=10$, 13, 16 and 18~$\mu K$ with line-types dot-dashed, dashed, dotted and solid respectively. We do not plot the remaining two mass functions because of the dearth of halos. Clearly the amplitude of the halo mass function increases with increasing quadrupole, as expected. Included in this plot are two simulations run with CDM initial conditions and normalised by $\sigma_8$=0.8 and 0.9 (dotted and solid red lines respectively). Surprisingly, we ran most of the simulations with MOND as well and found that the amplitude of the initial conditions makes virtually no difference to the final halo mass function (various blue lines). This is likely because the growth of tiny fluctuations is particularly fast in MOND and thus there is ample time from $z=200$ to $z\sim2$ for the lower normalisations to catch up to the point they get saturated, like speeding towards a traffic jam.

We also include in Fig~\ref{fig:quad} the mass functions for three different $\nu$ functions (both with $Q_{rms-PS}=17~\mu K$) which overlap with the other MOND mass functions. The three $\nu$ functions are parameterised as per \cite{famaey12} Eqs. 51 and 53 where

\beq
\protect\label{eqn:nubet}
\nu_{\beta}(y)=(1-e^{-y})^{-1/2}+\beta e^{-y}
\eeq
and
\beq
\protect\label{eqn:nualp}
\nu_{\alpha}(y)=\left[{1+(1+4y^{-\alpha})^{1/2}\over 2}\right]^{1/\alpha},
\eeq
where $\alpha=1$ is the so-called simple $\nu$-function and $\alpha=2$ is the standard $\nu$-function.

\begin{figure}
\includegraphics[angle=0,width=8.0cm]{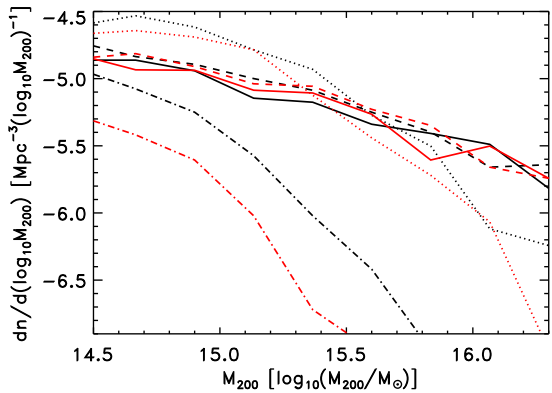}
\caption{Here we plot the comoving mass functions of two separate simulations both with 256~Mpc/h boxes, but different initial normalisations: one with  $Q_{rms-PS}=$17~$\mu K$ (black lines: table 1 - Q-17am) and the other with $Q_{rms-PS}=$10~$\mu K$ (red lines: table 1 - Q-10m). The different line-types correspond to the scale-factors at which the halo mass function was computed. The scale-factors were 1.0 (solid), 0.85 (dashed), 0.36 (dotted) and 0.21 (dot-dashed).  The two solid and dashed mass functions cannot easily be distinguished, however the dotted and dot-dashed mass functions clearly show the larger initial normalisation of the black lines.}
\label{fig:mfwz}
\end{figure}

To confirm that the growth of halos saturates in our MOND simulations, we have plotted in Fig~\ref{fig:mfwz} a number of simulated mass functions for two different sets of initial conditions: $Q_{rms-PS}=17\mu K$ (black lines) and $Q_{rms-PS}=10\mu K$ (red lines). The simulations use MOND and 256~Mpc/h boxes. The different line-types correspond to the scale-factors at which the comoving halo mass function was computed. The scale-factors were 1.0 (solid), 0.85 (dashed), 0.36 (dotted) and 0.21 (dot-dashed).  The two solid and dashed mass functions cannot easily be distinguished, however the dotted and dot-dashed mass functions clearly show the larger initial normalisation of the black lines. This means the two simulations began with very different density perturbations, but after $a=0.36$ they became indistinguishable.

The halos stop growing because they run out of matter to accrete. Initially, there are large reservoirs of matter surrounding the perturbations, but as this is used up, the accretion rate drops. In the Newtonian simulations with 256~Mpc/h boxes, the typical fraction of particles locked in halos (resolved by our simulations) of more than 200 particles is 0.1. For the MOND simulations described above, the fractions of particles locked in halos is larger. At scale-factors 0.21, 0.36, 0.49, 0.69, 0.85 and 1.0 the $Q_{rms-PS}=17 (10) \mu K$ simulations have 0.013 (0.005), 0.15 (0.10), 0.35 (0.31), 0.41 (0.38), 0.45 (0.42) and 0.47 (0.47) as fractions of particles locked in halos. Without dark energy, because of the logarithmic potential of MOND all mass would eventually become bound in halos. The acceleration of the universe at late times prevents this and only allows $\sim$50\% of matter to be bound to halos by $z=0$.

\begin{figure}
\includegraphics[angle=0,width=8.0cm]{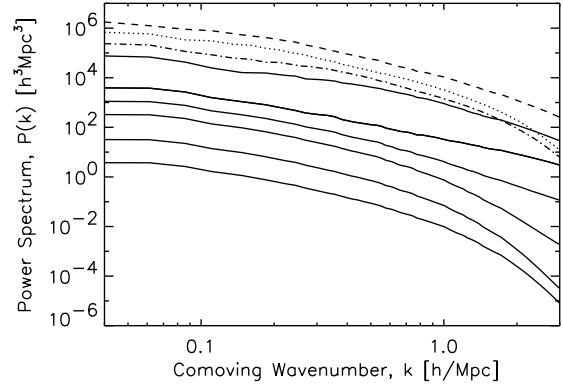}
\caption{Here we plot the particle power spectrum for our MOND simulation with $Q_{rms-PS}=$17~$\mu K$ (table 1 - Q-17am). From the bottom up, the solid lines are for different scale factors $7.45\times10^{-3}$, 0.016, 0.035, 0.05, 0.074, 0.11, 0.22. The dot-dashed, dotted and dashed lines are scale-factors 0.34, 0.53 and 1.0 respectively.}
\label{fig:kpk}
\end{figure}

Furthermore, we have plotted in Fig~\ref{fig:kpk} the power spectrum of the $Q_{rms-PS}=17 \mu K$ simulation at various redshifts for the wavenumbers probed by our code. The initial particle power spectrum correctly represents the analytical power on all scales.

\begin{figure}
\includegraphics[angle=0,width=8.0cm]{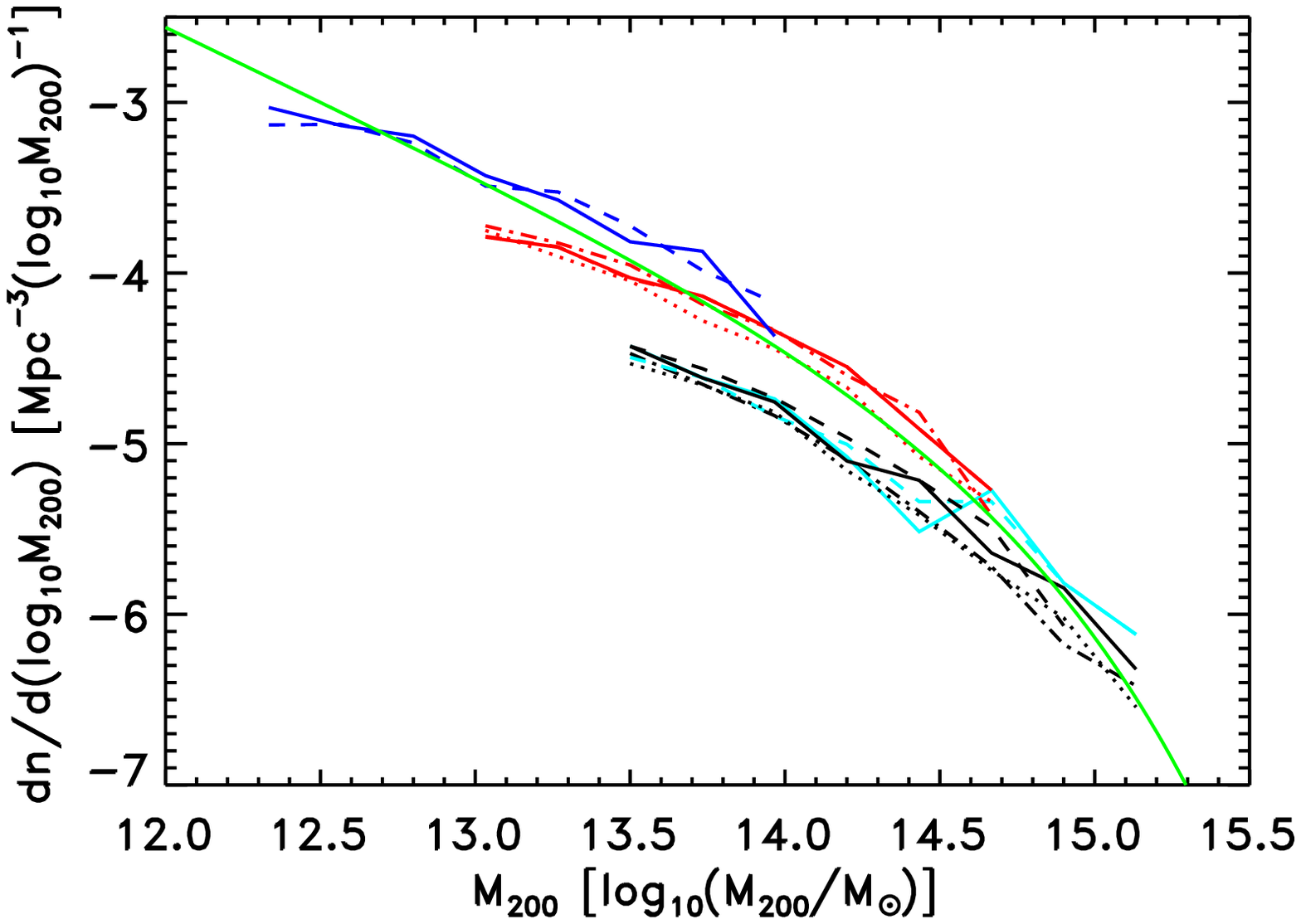}
\caption{Here we plot the Newtonian halo mass function of a $300~eV$ sterile neutrino using a series of box sizes. The blue, red and black coloured lines correspond to 64, 128, 256~$Mpc/h$ box lengths. The different line-types signify a different random realisation of the initial conditions with the same box size. In table 1 these correspond to simulations b64n-b128an (b*n and b*an use solid and dashed line-types respectively), Q-16n (solid) and Q-18n (dashed). We also plot one CDM simulations with $\sigma_8=0.8$ and a 128~$Mpc/h$ box (red dotted: cdm08n128), two with 256~$Mpc/h$ boxes (black dotted and dot-dashed: cdm08n and cdm08n1) and two 128~Mpc/h with only 129 cells per dimension (turquoise solid and dashed for CDM and 300~eV respectively: b128c129ncdm and b128c129n). The green solid line is the theoretical halo mass function (using the code of Reed et al. 2007) for CDM with $\sigma_8=0.8$ and our default parameters.}
\label{fig:300n}
\end{figure}

In summary it does not appear to be possible to form the correct halo mass function in standard MOND from any sterile neutrino initial conditions that grew from an initially Harrison-Zel'dovich power spectrum under GR until $z\sim200$. So if MOND is the correct description of gravitational dynamics on galaxy scales, then either the initial conditions are not as described above and yet conspire to produce the correct CMB angular power spectrum, or MOND does not affect the sterile neutrinos. This is important because although galaxies require MOND to form (and stably exist) without CDM, the clusters clearly do not require MOND at all and one should not ignore how well Newtonian gravity reproduces the cluster mass function (cf. Fig~\ref{fig:300n}).

At minimum, the CDM model gives the correct cluster scale halo mass function at $z=0$, whether some additional boost to gravity is required to form the clusters early enough has been discussed in the literature (\citealt{mullis05,bremer06,jee09,jee11,rosati09,brodwin10,brodwin12,foley11}). MOND has a double negative effect on the cluster mass function if it influences the sterile neutrinos. Not only does it facilitate more rapid growth and the formation of much larger and denser structures than in Newtonian gravity (meaning the MOND $M_p(r)$ is larger), but these more massive halos now have MOND gravity meaning their $M_{m,200}$ (Newtonian equivalent masses at $r_{200}$) are further enhanced, causing poorer agreement with the data. This result might suggest that if the MOND gravitational field is not produced by the sterile neutrinos (meaning only a Newtonian gravitational field is produced by them), but is only produced by the baryons, then it will have a positive influence on the halo mass function.

\section{Conclusion}
Here we tested the hypothesis that combining MOND with either an $11~eV$ sterile neutrino (as per the original proposal of \citealt{angus09}) or with a larger mass of sterile neutrino (up to $300~eV$) could produce the observed mass function of clusters of galaxies. We ran many cosmological simulations using the code of AD11 and found that the $11~eV$ sterile neutrino severely underpredicted the number of low mass clusters of galaxies and that it is completely ruled out. A 30-300~$eV$ sterile neutrino could produce the correct number of low mass clusters of galaxies, but greatly overproduced the number of high mass clusters of galaxies. We tested many proposed solutions: like reducing the normalisation of the initial conditions; at which redshift MOND switches on; variations on the expansion history; interpolating functions and found they were all ineffective. This means that, if MOND is the correct description of weak-field gravitational dynamics on galaxy scales, then either the whole cosmology and/or the initial conditions are not as described above (see, e.g., section 9.2 of \citealt{famaey12} for a discussion in the context of covariant MOND theories), and yet would conspire to produce the correct CMB angular power spectrum, or conversely MOND does not affect the sterile neutrinos in the same way as the baryons.
\section{acknowledgements} The authors greatly appreciate the efforts made by the referee that substantially improved the quality of the paper. GWA is a postdoctoral fellow of the FWO Vlaanderen (Belgium). Part of the research was carried out while GWA was a postdoctoral fellow supported by the Claude Leon Foundation. KJvdH's research is funded by the National Research Foundation of South Africa and in particular the NRF special award for Y-rated researchers is gratefully acknowledged. AD acknowledges partial support from the INFN grant PD51.

\end{document}